# Preparation of Monodisperse and Highly Coercive $L1_0$-FePt Nanoparticles Dispersible in Nonpolar Organic Solvents

Shinpei Yamamoto[*], Yasumasa Morimoto[†], Yoshinari Tamada, Y. K. Takahashi[‡], K. Hono[‡], Teruo Ono and Mikio Takano

*Institute for Chemical Research, Kyoto University, Uji 611-0011, Japan*

[†] : Present address: Toyobo Co. Ltd., 2-2-8, Dojimahama, Osaka, 530-8230, Japan

[‡] : National Institute for Materials Science (NIMS), 1-2-1, Sengen, Tsukuba, 305-0047, Japan

[*] : To whom correspondence should be addressed:
  e-mail : shinpei@msk.kuicr.kyoto-u.ac.jp
  phone number : +81-774-38-4718
  fax number : +81-774-38-3125

**Abstract:**

A method to prepare monodisperse and highly coercive $L1_0$-FePt nanoparticles which are dispersible in nonpolar organic solvents such as toluene, chloroform, and hexane was developed. By vigorously stirring the $SiO_2$-coated $L1_0$-FePt nanoparticles synthesized by the "$SiO_2$-nanoreactor" method (*Appl. Phys. Lett.* **2005**, *87,* 032503) in a mixture of an aqueous NaOH solution, chloroform, and hexadecyltrimethlyammonium bromide, the $SiO_2$ coating was dissolved off and bare FePt nanoparticles could be extracted to the chloroform phase without degrading their magnetic properties. These particles showed a tendency to form a close packed lattice on slow drying of the chloroform-based solution. The present success may promote the practical application to ultra-high density magnetic recording and also may open the door to providing these particles with new physical and/or chemical functions.

**Introduction**

The FePt alloy with the $L1_0$ structure possesses a very high uniaxial magnetic anisotropy of ca. $6 \times 10^6$ J/m$^3$, which is more than ten times as high as that of the currently utilized CoCr-based alloys.[1] Superparamagntic fluctuation of the room temperature magnetization can thus be suppressed even for particles of 3 nm in diameter, making an appropriate array of these nanoparticles to be a promising candidate for future ultra-high density magnetic recording media of >1 Tbit/inch$^2$.[2-9] The most basic requisite for the practical use would be the formation and the fixation of an array on a substrate with the magnetic easy axis, *i.e.,* the tetragonal *c* axis, oriented normal to the substrate surface. Such a close packed triangular array structure may be formed through self-organization under an external magnetic field if the $L1_0$-FePt nanoparticles could be dispersed in a polymer binder. However, it has been experimentally difficult because of several drawbacks like insufficient Fe/Pt ordering, nonuniformity in particle size and shape, and lack of solubility in nonpolar organic solvents that dissolve a polymer binder.[4c, 5b,c]

Recently we have succeeded in preparing monodisperse and highly coercive $L1_0$–FePt nanoparticles which are dispersible in water by the strategic "SiO$_2$-nanoreactor" method.[10] These particles are well-crystallized and have a room temperature coercivity of 19 kOe in spite of their diameter of only 6.5 nm. Also the particle moment could be highly oriented. Here, we report a method to render these $L1_0$-FePt nanoparticles dispersible in nonpolar organic solvents, *e.g.,* toluene, chloroform, and hexane. We mean that we succeeded in removing the SiO$_2$-shell and extracting the $L1_0$-FePt cores in these nonpolar organic solvents without degrading their high coercivity.

**Experimental Section**

   **Chemicals.** Octyl ether (>95 %, Tokyo Kasei), platinum(II)-acetylacetonate (first grade, Wako Pure Chemicals), teraethyl orthosilicate (special grade, Wako Pure Chemicals), oleic acid (first grade, Wako Pure Chemicals), oleyl amine (97 %, Acros Organics), pentacarbonyl iron (99.99 %, Aldrich), 1,2-hexadecandiol (90 %, Aldrich), and hexadecyltrimethlyammonium bromide (HTAB, GR grade, Nacalai Tesque) were used as received.

   **Preparation of the $SiO_2$-coated $L1_0$-FePt nanoparticles.** The $SiO_2$-coated $L1_0$-FePt nanoparticles were prepared according to the procedures described elsewhere.[10a] In brief, face-centered cubic (fcc) FePt nanoparticles with an average particle size of 6.4 nm were prepared first according to the method reported previously.[3a] They were made water-soluble by encapsulation in HTAB using an oil-in-water microemulsion technique,[11] and their hydrophilic surface was subsequently coated by $SiO_2$. The $SiO_2$-coated fcc-FePt nanoparticles were then heated in a tube furnace (Lindberg, KTF045) at 900 °C for 1 h in a stream of $H_2$(5 %)/Ar(95 %) for the transformation to the $L1_0$ structure. The previous work proved that this treatment provides well-crystallized and isolated $L1_0$ particles of 6.5 nm in diameter.[10a,b]

   **Preparation of the $L1_0$-FePt nanoparticles dispersible in nonpolar organic solvents.** The $L1_0$-FePt nanoparticles dispersible in nonpolar organic solvents were prepared via an aqueous/organic biphasic reaction using HTAB as a phase-transfer reagent. In a typical run, finely ground $SiO_2$-coated $L1_0$-FePt nanoparticles (0.03 g), aqueous NaOH solution (4 M, 3 g), chloroform (5 g), and HTAB (0.5 g) were mixed and vigorously stirred for 24 hr at room temperature. In this process the $SiO_2$ layer was dissolved off and bare $L1_0$-FePt nanoparticles were extracted to the chloroform phase. These were collected by centrifugation after adding ethanol to decrease the solubility and were again dispersed in chloroform containing oleic acid and oleyl amine. Undesirable precipitates were removed by centrifugation, and the $L1_0$-FePt nanoparticles left in the supernatant were collected by centrifugation after adding ethanol. Thus-obtained powder was redispersed in chloroform containing oleic acid and oleyl amine and stored at room temperature.

**Characterization.** Transmission electron microscopy (TEM) observations were performed using a Philips CM 200 TEM and a JEOL JEM-4000EX. TEM specimens were prepared by dropping the particle-containing original solution on a carbon-coated copper grid. The particle composition was determined by means of Energy-dispersive X-ray analysis (EDX) on a JEOL JED 2140. Fourier transform infrared (FTIR) spectra were collected on a BioRad FTS-6000. The samples were mixed with KBr and compressed into pellets. Magnetic properties were measured with a SQUID magnetometer (Quantum Design MPMS-XL). For the EDX and FTIR measurements, the nanoparticles were precipitated from the solution by adding ethanol and then dried.

**Results and Discussions**

In the low-magnification TEM image in Fig. 1a we can see that the $L1_0$-FePt nanoparticles are dispersed free of aggregation. The average particle size has been estimated to be 6.5 nm, which is the same as that of the $SiO_2$-coated FePt particles. Their composition was determined to be $Fe_{55}Pt_{45}$, which is again the same as that of the original fcc-nanoparticles. Inset shows a bottle of the chloroform solution of the $L1_0$-FePt nanoparticles. Dispersability in other nonpolar organic solvents including toluene, hexane, and others was confirmed. By adding small amounts of oleic acid and oleyl amine to them, we could keep the solutions stable for months, free from any precipitation.

Thanks to their uniformity in size and also to their nearly spherical shape, the $L1_0$-FePt nanoparticles tended to form 2- and 3-dimensional close-packed superlattices through spontaneous self-organization on slow drying as can be seen in Fig. 1b. The inter-particle distance is ~10 nm, from which the thickness of the coating layer made of oleic acid and oleyl amine has been estimated to be ~1.8 nm.

Figure 2 shows a typical selective area electron diffraction (SAED) pattern. Note the appearance of the (110) ring characteristic of the $L1_0$-structure with $d_{110} = 0.27$ nm.[12] The high-resolution TEM (HRTEM) image in Fig. 3 clearly shows that the $L1_0$-FePt nanoparticles are single crystalline with the (110)-superlattice contrast clearly seen up to the surface end. The particle is facetted on {111} and {011} planes.

These solutions could dissolve a variety of polymers such as poly(methyl methacrylate) and polystyrene, from which thin polymer films containing uniformly dispersed $L1_0$-FePt nanoparticles could be easily formed on various kinds of substrates. Their chemical and physical properties are currently under investigation.

Here, we discuss various steps of the extraction process. The first step is the dissolution of the $SiO_2$-coating. It was amorphous in nature even after the treatment at 900 °C in flowing $H_2$(5 %)/Ar(95 %) atmosphere as found in our previous work.[10a] As well known $SiO_2$ depolymerizes under strong basic conditions.[13] In the previous work we dissolved it just by suspending in a tetramethylammonium hydroxide solution (10 wt% aqueous solution), while in the present process basicity was provided by the aqueous NaOH solution. The bared $L1_0$-FePt nanoparticles thus obtained were then extracted to the chloroform phase with the aid of HTAB. The mechanism contains most probably the formation of a double-layer such as schematically drawn in Fig. 4 which consists of negatively charged hydroxide ions adsorbed to the surface of the $L1_0$-FePt nanoparticles and counteractions of hexadecyltrimethlyammonium (HTA). A small amount of water might also be contained as suggested in the same figure. The FTIR spectrum of the $L1_0$-FePt nanoparticles just after the extraction to the chloroform phase has peaks assigned to HTA (Fig. S1 and Fig. S2 (B)), which supports this double-layer view. These $L1_0$-FePt nanoparticles could favor being in the chloroform phase rather than in the aqueous phase owing to the hydrophobic character of the long alkyl chains of HTA. In the extraction process, the countercations are supposed to play two important roles: one is to form the double-layer while in the aqueous phase,[14] and the other is to promote dispersion in the chloroform phase. In other words, they should have a suitable degree of amphiphilicity for successful extraction. Actually, the use of a more hydrophobic cation like trioctylmethylammonium in place of HTA and also the use of a more hydrophilic cation of tetramethylammonium resulted in very limited success. The moderately hydrophobic and hydrophilic character of HTA must have been the key to the successful extraction.

    Figure 5 shows FTIR spectra of certain samples. The spectrum of the $SiO_2$-coated $L1_0$-FePt nanoparticles ((a), Fig. S2 (A)) consists of essentially two peaks around 1100 and 800 $cm^{-1}$ assigned to amorphous $SiO_2$. On the other hand, the spectrum of the $L1_0$-FePt nanoparticles after the purification process ((b), Fig. S2 (C)) has no peaks of $SiO_2$-origin but has new peaks arising from the organic stabilizers. Inset shows the spectrum magnified in the 2750-3030 $cm^{-1}$ region. The observed peaks are assigned to the oleyl group,[15] though the spectrum taken just after the extraction to the chloroform

phase showed peaks from HTA. Replacement of HTA by oleic acid and oleyl amine must have taken place in the subsequent purification steps (Fig. S2).

Magnetic properties of dried and chloroform-dispersed $L1_0$-FePt nanoparticles studied through magnetization measurements using a SQUID magnetometer were essentially the same as those reported in the previous paper.[10] Shown in Fig. 6 for example is the hysteresis loop of the chloroform-dispersed nanoparticles measured after field-cooling down to 150 K at 10 kOe.[16] At this temperature, the matrix solution was frozen, any motion of the FePt nanoparticles thus being suppressed. Here, $M_s$ represents the saturation magnetization.[17] It is notable that the remanent magnetization ($M_r$: magnetization at $H$=0 kOe) is almost equal to the saturation magnetization as expected from an anisotropic hard magnet. On reversing the applied field, the magnetization shows a small and quick drop to $M / M_s$ = 0.83 and then decreases very slowly down to $M / M_s$ = -0.63, not to -1 as expected. This indicates that the reversal of magnetization did not finish completely because the maximum field available was not large enough in comparison with the coercivity of ~-40 kOe. The initial small drop mentioned above is attributed to the presence of a small amount of untransformed fcc-FePt nanoparticles as suggested from Mössbauer spectroscopic measurements on the $SiO_2$-coated samples.[10c] We note here that it is beyond the scope of the present work to measure the degree of the alignment of particle moment as a function of temperature, applied field, and particle size for comparison with a very recent theoretical calculation by Bain *et al*.[18].

**Conclusions**

We have developed a method to prepare monodisperse and magnetically highly coercive $L1_0$-FePt nanoparticles dispersible in nonpolar organic solvents like toluene, chloroform, hexane. We believe that the solubility in nonpolar organic solvents will allow us to attach new functionalities to the surface of the $L1_0$-FePt nanoparticles through a variety of organic and organometallic chemical reactions. Such $L1_0$-FePt nanoparticles will be of great potentiality not only for ultra-high density magnetic recording but also for magnetochemical and magnetophysical surface functionalities.

**Acknowledgement**

The authors express their thanks to the Ministry of Education, Culture, Sports, Science and Technology, Japan, for Grants-in-Aid No. 12CE2005, for COE Research on Elements Science and No. 14204070, and for 21COE on Kyoto Alliance for Chemistry. S.Y. expresses his thanks to C. E. McNamee of Kyoto University for beneficial discussion.

**Figure captions**

**Figure 1**: (a) Low-magnification TEM images of the $L1_0$-FePt nanoparticles. (b) Tendency to form ordered superlattice structure. Inset shows an image of the chloroform solution containing the $L1_0$-FePt nanoparticles.

**Figure 2**: SAED pattern of the $L1_0$-FePt nanoparticles.

**Figure 3**: HRTEM image of the $L1_0$-FePt nanoparticle.

**Figure 4**: Schematic illustration of the double-layer.

**Figure 5**: FTIR spectra of the $L1_0$-FePt nanoparticles at the various preparative steps: (a) the $SiO_2$-coated state and (b) after the purification step. Inset shows the FTIR spectrum of the $L1_0$-FePt nanoparticles after the purification step in the 2750-3030 cm$^{-1}$ region. The arrows indicate the characteristic absorption peaks of the oleyl group.[15]

**Figure 6**: Hysteresis loop of the chloroform-dispersed $L1_0$-FePt nanoparticles measured at 150 K after cooling in an applied magnetic field of 10 kOe.

**Figure S1**: FTIR spectrum of the $L1_0$-FePt nanoparticles just after the extraction to the chloroform phase in the 2750-3030 cm$^{-1}$ region. The arrows indicate the characteristic absorption peaks of the alkyl groups of HTA: The peaks around 2850 and 2920 cm$^{-1}$ are assigned to the symmetric and asymmetric $CH_2$ stretching modes. The peak around 2870 cm$^{-1}$ is assigned to the symmetric $CH_3$ stretching mode. The broad peak around 2960 cm$^{-1}$ is assigned to the two unresolved asymmetric $CH_3$ stretching modes. The peak around 3010 cm$^{-1}$ is assigned to the asymmetric $CH_3$ stretching mode of the trimethylammonium headgroup.

**Figure S2**: Schematic representation of the extraction and purification processes.

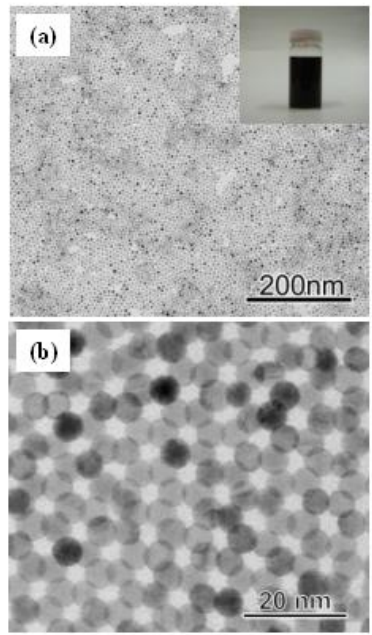

Figure 1

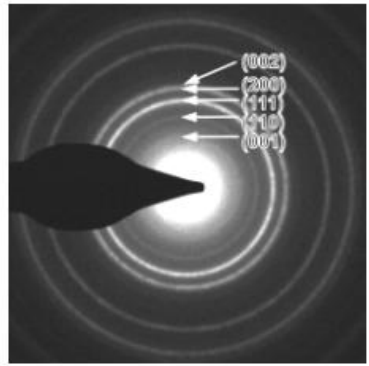

Figure 2

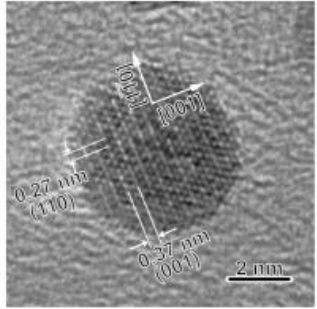

Figure 3

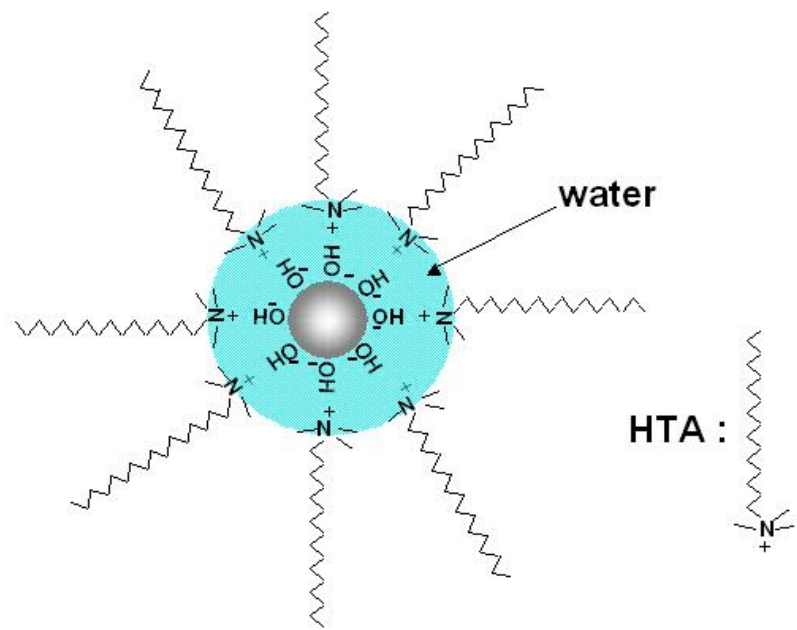

Figure 4

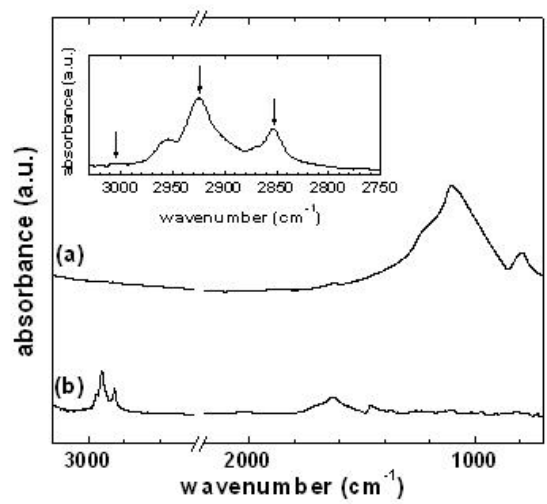

Figure 5

Figure 6

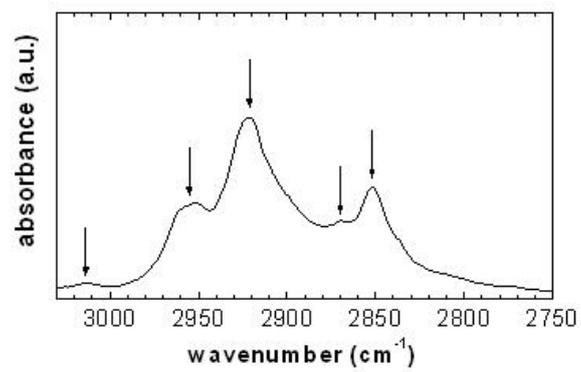

Figure S1

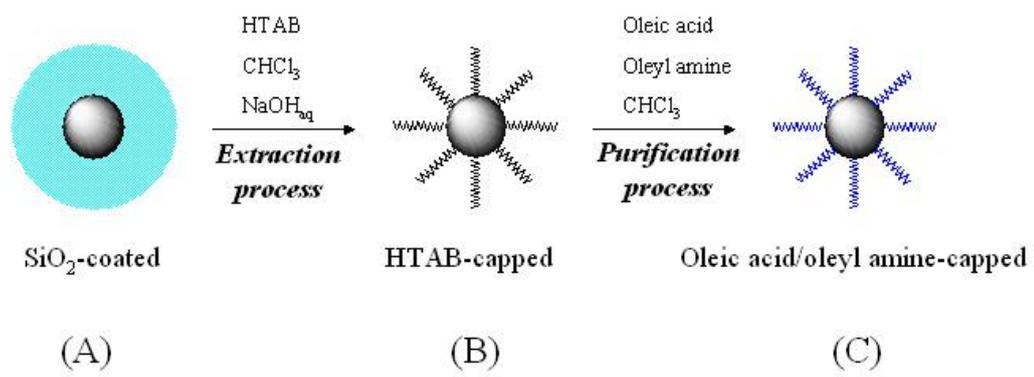

Figure S2